\documentclass[doublecol]{epl2}
\usepackage{amsmath}
\usepackage{subcaption}
\usepackage{epstopdf}

\def\rs{{\bf{r}_s}}
\def\r{{\bf{r}}}

\def\n{{ \bf n}}

\def\v{{ \bf v}}
\def \G{{G(\rs,\r)}}

\def\bnabla{{ \bf \nabla}}

\def\l[{{[\![}}
\def\rt]{{]\!]}}
\def\zsh{z_\mathrm{shift}}

\title{Translation-deformation coupling effects on the Rayleigh instability of an electrodynamically levitated charged droplet }

\shorttitle{Translation-deformation coupling effects on the Rayleigh instability} 

\author{Neha Gawande\inst{1}, Y. S. Mayya \and Rochish Thaokar\inst{1}}
\shortauthor{N.Gawande, Y. S. Mayya and R. Thaokar }

\institute{                    
  \inst{1} Department of Chemical Engineering, Indian  Institute of  Technology Bombay, Mumbai-400076.\\
}

\abstract{
The breakup pathway of the Rayleigh fission process observed experimentally using high-speed imaging of a charged drop levitated in an AC quadrupole trap is shown to undergo asymmetric breakup by ejecting a jet in the upward direction ((i.e., opposite to the direction of gravity)). To explain this typical experimental observation, we carry out numerical calculations based on the boundary element method considering inertial droplets levitated electrodynamically using quadrupole electric fields. The simulations show that the gravity-induced downward shift in the equilibrium position of the drop in the trap causes significant, large-amplitude shape oscillations superimposed over the center-of-mass oscillations of the drop. An important observation here is that the shape oscillations due to the applied quadrupole fields, result in sufficient deformations that act as triggers for the onset of the instability below the Rayleigh limit, thereby admitting a sub-critical instability. The center-of-mass oscillations of the droplet within the trap, which follow the applied frequency, are out of phase with the applied AC signal. Thus the combined effect of shape deformations and dynamic position of the drop leads to an asymmetric breakup such that the Rayleigh fission occurs upwards via the ejection of a jet at the north-pole of the deformed drop.
}
\pacs{nn.mm.xx}{First pacs description}
\pacs{nn.mm.xx}{Second pacs description}
\pacs{nn.mm.xx}{Third pacs description}

\begin{document}

\maketitle

An isolated charged droplet is known to become unstable when the repulsive Coulombic force on account of its charge, just overcomes the stabilizing surface tension force of the droplet. This instability is popularly known as the Rayleigh instability of a charged droplet and the critical value of charge at which the instability sets in is given by the expression, $Q_R = 8\pi \sqrt{\gamma \epsilon_e a^3}$ where $a$ is the radius of the drop, $\epsilon_e$ is the permittivity of the external medium (air) and $\gamma$ represents the interfacial tension \cite{rayleigh1882}. Thus when the charge on the droplet increases beyond $Q_R$, it cannot sustain its stable spherical shape and exhibits deformation with time, ultimately leading to its breakup. To demonstrate the Rayleigh instability of a charged droplet and its subsequent deformation pathway with time, experiments were carried out by electrodynamically levitating a micron sized charged droplet in a quadrupole trap \cite{duft02},\cite{duft03}. In these experiments, a droplet is levitated perfectly at the center of the quadrupole field by using superimposed DC voltage on the AC field to balance the gravitational force acting on the droplet. Thus a critically charged droplet exhibits a sequential deformation from the original spherical shape to an elongated prolate spheroid, eventually forming symmetric conical tips from which two jets are ejected out in the opposite directions \cite{duft03}, \cite{giglio08}. This symmetric breakup, observed in experiments \cite{duft02},\cite{duft03}, is well predicted by numerical calculations \cite{betelu2006, gawande2017, gawande2020}.

The symmetrical jet ejection of a droplet perfectly levitated at the center of the quadrupole trap, may not correspond to practical situations such as electrosprays, wherein unbalanced external forces such as gravity or external electric field introduce asymmetry in the drop shape due to different local hydrostatic pressure and asymmetric charge accumulation due to non-uniform fields \cite{feng1991three}. The broken symmetry can not only impact the pathway of drop deformation and breakup but can also modify the critical limit of charge at which the drop is rendered unstable. 

Theoretical studies using both analytical approach \cite{adornato83}, \cite{natarajan1987}, \cite{pelekasis1990equilibrium}, \cite{tsamopoulos85} and numerical calculations \cite{basaran1989}, \cite{pelekasis1995dynamics}, \cite{das15}, indicate several interesting aspects: (i) The symmetric instability of an isolated charged droplet is subcritical with respect to "finite-amplitude" prolate spheroidal perturbations, in fact a perfect transcritical bifurcation is admitted at the critical Rayleigh charge ($Q_R$). The finite amplitude perturbation was always imposed, in an artificial manner, as an initial condition. (ii) The numerical analysis of charged drops under an externally applied DC uniform electric field showed that the drops could become unstable in the sub-Rayleigh limit when the applied field is above a critical value \cite{basaran1989}. The finite amplitude perturbations, supplied as an initial condition, in this case are amplified by the applied uniform DC field, thereby admitting an imperfect transcritical bifurcation. (iii) A similar analysis was done recently considering highly charged drops under DC quadrupole electric field, and the effect of the field on Rayleigh critical charge was explained via a detailed bifurcation diagram \cite{das15}, that also admitted a transcritical bifurcation. In this work, it was assumed that a charged droplet is quasi-statically levitated at the center of the quadrupole field, and the absence of electrophoresis led to symmetric breakup of the charged droplet similar to the experimental observation of Duft \etal \cite{duft03}.

The above numerical studies involved either a DC uniform or a DC quadrupolar field. In experiments though, systematic study on levitated charged droplets is carried out using AC quadrupolar fields \cite{duft02}\cite{giglio08}. Herein, charged droplets are levitated electrodynamically in the quadrupole trap using AC electric fields and exhibit a continuous center of mass (COM) oscillations across its equilibrium position in the trap. Thus it is necessary to consider the translational motion of a droplet in the presence of an applied quadrupole field to accurately model the system to study the breakup of the levitated charged droplet. Recently, the experiments carried out on such electrodynamically levitated charged droplets have shown that in the presence of gravitational force acting on the droplet, the equilibrium position of the drop is shifted from the center of the quadrupole trap in the downward direction \cite{singh2019effect}\cite{singh2018pof}. This shift is responsible for sustained oscillations of the COM motion, simultaneously causing significant, finite amplitude deformation of the droplet. Moreover, the off-center position of the charged droplet introduces an asymmetry in the drop shape deformation. All these ingredients, as will be demonstrated in this work, ultimately lead to an asymmetric, sub-critical instability. 

In our previous work, we attempted to address and understand this problem in the Stokes flow limit \cite{pre_paper}. The goal here was to predict the upward breakup observed in most experiments. Since the drop breakup completes in $1/4^{th}$ of the AC cycle, the equilibrium position of the droplet in the quadrupole trap (in terms of $\zsh$, where $\zsh$ represents distance of the droplet center from the center of the quadrupole field) was obtained by comparing experimental observation with the solution of the modified Mathieu equation, and the BEM simulations were carried out at this position, with a perturbed drop shape as an initial condition (obtained from the experiments), after which the drop continuously deforms leading to breakup. The droplet COM position could not be simulated since in the Stokes flow limit, quasi-static condition is assumed and the inertial forces are neglected. The directionality of the asymmetric breakup of the charged droplet was observed to critically depend on the position of the drop in the trap and the magnitude of perturbation attained by the drop at the end of the oscillation phase \cite{pre_paper}. 

An asymptotic theory suggests that the surface oscillations of a charged droplet in the quadrupole field can be explained using potential flow equations. Thus to accurately capture the effect of COM motion of the levitated charged droplet on its simultaneous shape deformation that leads to breakup, the numerical simulations are needed to be carried out in the inertial timescales. Towards this, we developed numerical codes based on the boundary integral formulation in the potential flow limit. Here the system considered is of a liquid droplet levitated in the air. As the density of air ($\rho_e=1.2$) is three orders of magnitude lower than the ethylene glycol drop ($\rho_i=1097$), we have restricted our analysis to a liquid droplet suspended in the dynamically inactive external medium. Thus the boundary integral equation for velocity potential inside the drop is solved to study the oscillation dynamics of the drop. Here we use the integral formulation based on the generalized vortex method, developed by Baker et al. (1982) \cite{baker1982generalized} and applied by Lundgren and Mansour (1988) \cite{lundgren1988oscillations} for an inviscid liquid drop in vacuum or gas with negligible density.  

\begin{figure}
	\begin{center}	
		\includegraphics[width=0.35\textwidth]{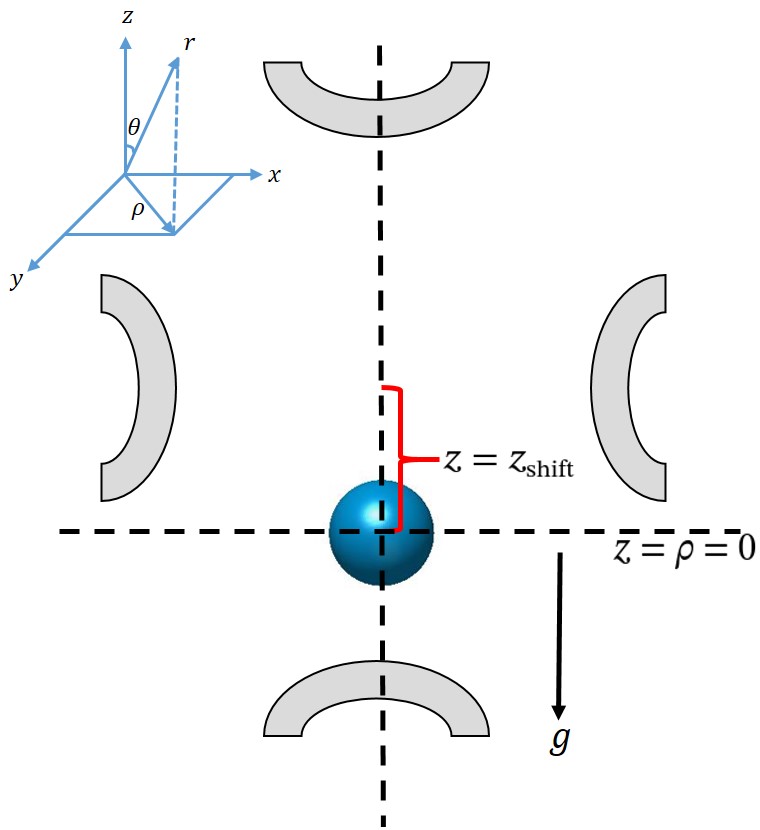}
		\caption{Schematic of setup used in the numerical simulations indicating position of the drop in the quadrupole trap with four electrodes.}
		\label{fig:numsetup}
	\end{center}
\end{figure}

Since the electrical conductivity of ethylene glycol drop used in the experiments is high \cite{pre_paper}, the relative charge relaxation timescale ($\tau_e=\epsilon_i/\sigma_i$, where $\epsilon_i$ and $\sigma_i$ represent the permittivity and conductivity of the drop respectively) is much smaller than the capillary timescale $(\tau_c=\sqrt{\rho_i a^3/\gamma})$. For typical experimental parameters, the ratio of two timescales, known as Saville number, $Sa=t_e/t_c \sim O(10^{-4}$). Thus in this study, the surface charge dynamics are neglected, and the drop is assumed to be perfectly conducting drop with a net electric charge $Q$ distributed uniformly on the drop surface and the drop is suspended in a perfectly dielectric surrounding medium with a permittivity, $\epsilon_e$. The equations can be suitably non-dimensionalised by characteristic scales as; the time by the inertial timescale $\tau_c$, the pressure by $\gamma/a$; while the charge and electric fields are scaled by $\sqrt{a^3 \gamma \epsilon_e}$ and $\sqrt{\gamma \epsilon_e/a}$ respectively such that the non dimensional Rayleigh charge is $Q=8\pi$. 

The non-dimensional electric potential on the surface of the drop is denoted by (${\phi}$ ) and is assumed to follow,
\begin{equation}
    {\nabla}^2{\phi}=0.
\end{equation}
The electric field is thus expressed as $\bf{E}=-{\nabla}{\phi}$.

The flow is assumed to be irrotational inside the drop; thus the velocity is given by the gradient of the velocity potential ${\v}={\bnabla}{\psi}$ such that velocity potential follows the Laplace equation,

\begin{equation}
 \bnabla^2\psi=0   
\end{equation}

\begin{figure}
	\begin{center}	
		\includegraphics[width=0.48\textwidth]{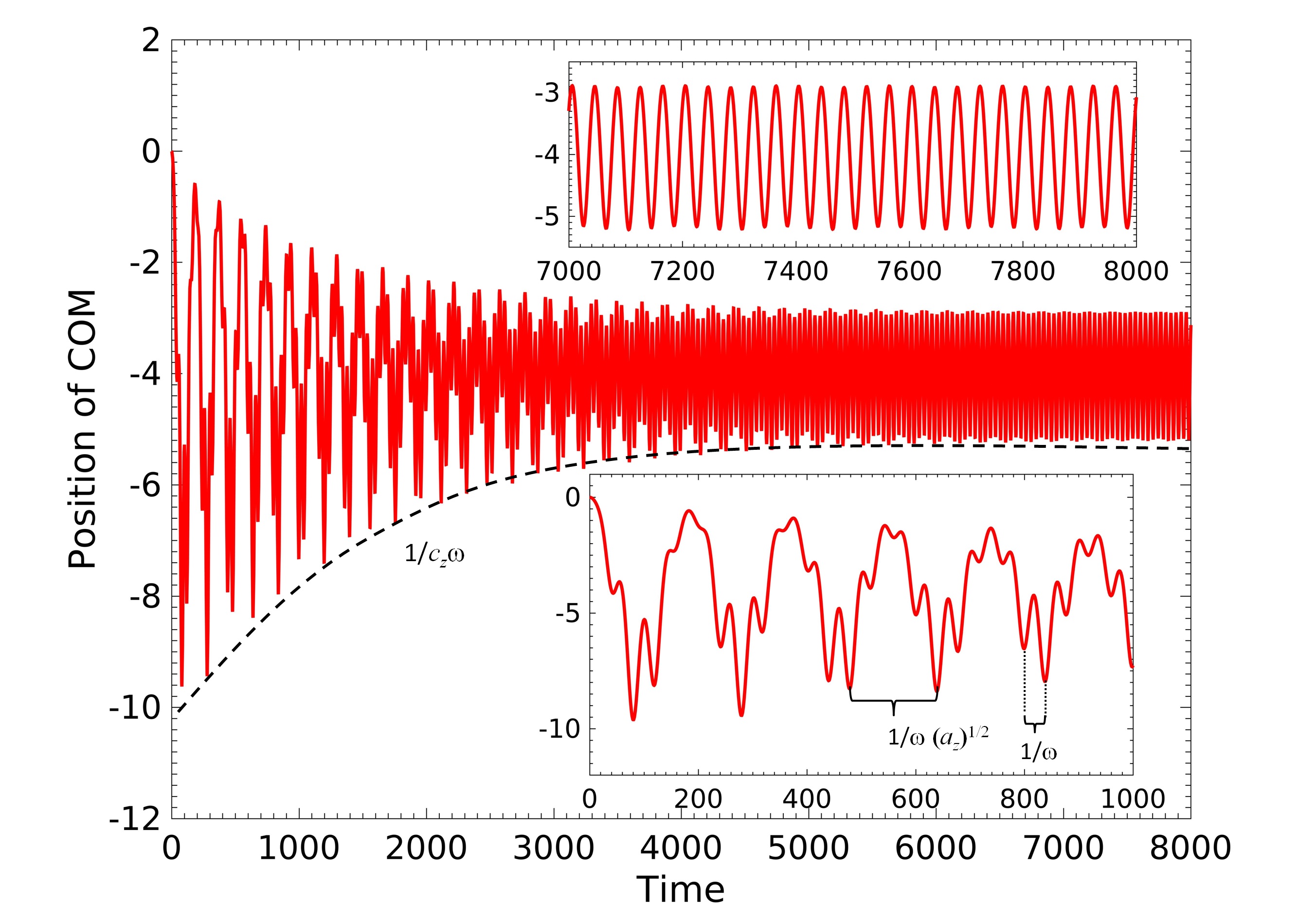}
		\caption[Oscillations exhibited by COM of the levitated droplet indicating three different timescales.]{Oscillations exhibited by COM of the levitated droplet indicating damping over the timescale of $1/c_z\omega$ (indicated by black dotted line). The two timescales due to applied frequency $1/\omega$ and secular frequency $1/\sqrt{a_z}\omega$ are shown in the bottom inset while top inset shows the zoomed in COM motion at steady state.}
		\label{fig:com_damping}
	\end{center}
\end{figure}

\begin{figure}
	\centering
	\includegraphics[width=0.3\textwidth]{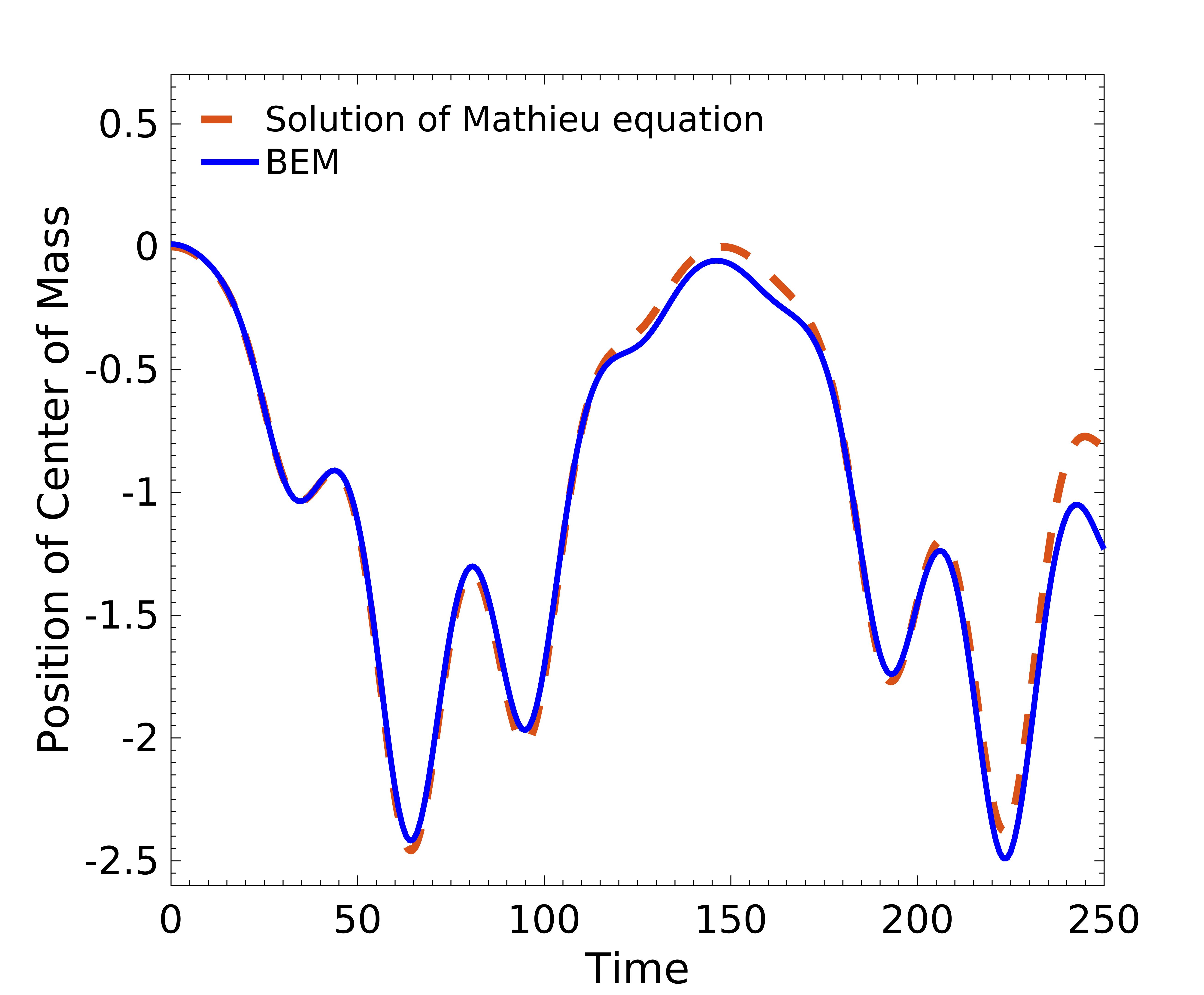}
	\caption[Temporal evolution of the COM motion of the droplet obtained from BEM results compared with the numerical solution of the modified Mathieu equation.]{Temporal evolution of the COM motion of the droplet obtained from BEM results compared with the numerical solution of the modified Mathieu equation (equation \ref{eqn:mathieu_break}) in the absence of damping. Parameters used are: $Q=6\pi$, $Bo=0.0015$, $Ca_\Lambda=0.0012$ and $f=0.03$.}
	\label{fig:val_mathieu}
\end{figure}

\subsection{Integral equation for electric potential} 
 For the case of perfectly conducting charged drop in AC quadrupole electric fields, the integral equation for the electric potential is given by,
 \begin{equation}
 \phi({\bf r_s})=\phi_\infty({\bf r_s})+\int \G E_{n} dA({\bf r})
 \end{equation}
 where, $\G=1/4\pi (\r-\rs)$ and ${\bf{r}}$ and ${\bf r_s}$ are the position vectors on the surface of the drop while $\phi_\infty$ is the applied electric potential which can be written as,
 \begin{equation}
 \phi_\infty(\rho,z)=\sqrt{Ca_\Lambda}\zeta(t)[(z-\zsh)^2-0.5\rho^2]
 \label{eqn:simu_cal}
 \end{equation} 
 where, $Ca_\Lambda$ is the non-dimensional intensity of quadrupole field and $\zeta(t)$ is any time varying function with frequency $\omega=2\pi f$. $(z-\zsh)$ accounts for the shifted position of the droplet from the center of the quadrupole trap.
 The unknown potential $\phi({\bf r_s})$ is constant on the surface of the drop, and is determined by the condition of conservation of charge, $$\int E_{ne} ({\bf r}) dA({\bf r})=Q$$ where $E_n$ is the outward directed normal electric field acting on the drop surface.
 
 \subsection{Integral equation for velocity potential}
 According to classical potential flow theory the velocity potential ($\psi$) of an irrotational flow can be expressed as a surface distribution of dipole density per unit area, $\vartheta(\rs)$ at the source point $\rs$ \cite{baker1982generalized}. This is also known as double layer potential representation and can be written as. 
 \begin{equation}
 \psi({\rs})=\frac{1}{2}\vartheta(\rs)+\int \vartheta(\r) (\n \cdot \bnabla \G) dA 
 \label{eqn:psi}
 \end{equation}
 
 For known values of scalar velocity potential ($\psi$) on the surface of the drop, equation \ref{eqn:psi} becomes a Fredholm integral equation of second kind for dipole strengths $\vartheta(\rs)$. This dipole density distribution is then used to obtain vector potential using integral equation \cite{lundgren1988oscillations},  
 \begin{equation}
 {\bf \mathcal{A}}(\rs)=-\int \vartheta(\r) (\n \times \bnabla_s \G) dA(\r)
 \label{eqn:vectorPot2}
 \end{equation}  
 where, ${\bf \mathcal{A}}(\rs)$ is the vector velocity potential and is related to normal component of velocity through,  
 \begin{equation}
 \v \cdot \n=(\n \times \bnabla) \cdot {\bf \mathcal{A}}
 \end{equation} 
 
To desingularize the integrals, we follow \cite{lundgren1988oscillations}, and the regularization of the kernels in equations \ref{eqn:psi} and \ref{eqn:vectorPot2} gives the final integral equations for velocity potentials as,
 \begin{align}
 \psi({\rs})&=\vartheta(\rs)+\int [\vartheta(\r)-\vartheta(\rs)] (\n \cdot \bnabla \G) dA(\r) \\
 {\bf \mathcal{A}}(\rs)&=-\int [\vartheta(\r)-\vartheta(\rs)] (\n \times \bnabla_s \G) dA(\r)
 \end{align}

where, $\vartheta$ represents dipole density distribution per unit area. The detailed expansions of $(\n \cdot \bnabla \G)$ and $(\n \times \bnabla_s \G)$ are given in our previous work \cite{singh2020influence}. Once the surface velocity is known, the velocity potential is updated using non-dimensional unsteady Bernoulli equation,
\begin{equation}
\frac{D \psi}{Dt}=\frac{1}{2} \v \cdot \v-\kappa-Bo \thinspace z+\frac{1}{2}E_n^2
\end{equation} 

where, $Bo=\rho_i g a^2/\gamma$ is the gravitational bond number and if $\n$ is the outward unit normal $\bnabla_s \cdot \n$ gives the curvature of the drop denoted by $\kappa$.

The simulations in the potential flow limit capture both the COM motion and large amplitude surface oscillations exhibited by a dynamically levitated charged droplet in the presence of gravity. It is observed that, in the presence of gravity, the deformation dynamics of the drop is significantly altered by its corresponding COM motion in the quadrupole trap, even in the small deformation limit \cite{singh2018pof}. Thus in this work, we extend our numerical analysis to understand the dynamical effect of COM motion of the levitated charged droplet on its breakup behavior.

\begin{figure}[tb]
	\begin{center}	
		\includegraphics[width=0.48\textwidth]{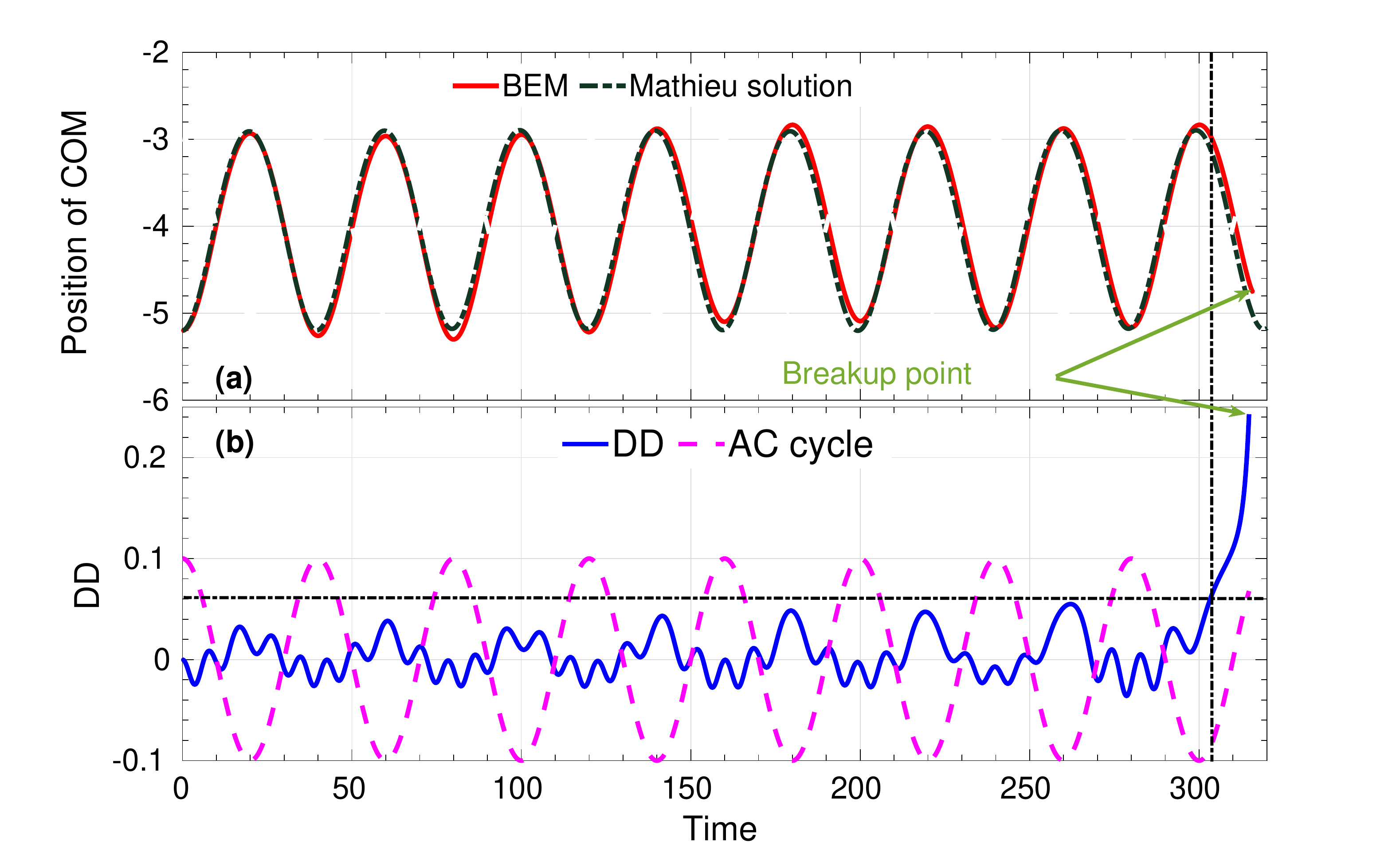}
		\caption[Position of COM and degree of surface deformation of a charged drop as a function of time with the corresponding applied AC field of the form $\cos (\omega t)$.]{(a)The position of COM as a function of time obtained from BEM and solving Mathieu equation.(b) Degree of deformation as a function of time with the corresponding applied AC field ($\cos (\omega t)$). Note that the amplitude of applied AC signal is scaled for better visualization. The intersection point of vertical and horizontal dash-dotted black line indicates the critical point of maximum deformation attained by the droplet after which it grows leading to breakup.}
		\label{fig:com_ddvstime}
	\end{center}
\end{figure}

In the experiments, the droplet is observed to oscillate about its equilibrium position with the applied frequency before the instability sets in due to a high surface charge near to its Rayleigh limit. The COM motion stability in the quadrupole field of an un-deformed spherical droplet is described by the solution of modified Mathieu equation which is written in current non-dimensional parameters as,
\begin{equation}
z''(t)+c_z z'(t)-a_z z(t) \cos(\omega t)+ Bo=0,
\label{eqn:mathieu_break}
\end{equation}
where, $a_z=(2 Q \Lambda_0 \tau_c^2)/m$, $c_z=6\pi\mu_e a\tau_c$ and $\omega=2\pi f$ with $f=\tilde{f}\tau_c$ as the non-dimensional applied frequency.
Since the numerical simulations presented here are carried out in the potential flow limit, without the inclusion of viscosity, the COM motion of the charged droplet exhibits conserved oscillations consisting of both applied and secular frequencies. The solution of equation \ref{eqn:mathieu_break} indicates that due to the presence of viscous drag (with damping coefficient $c_z$), the secular oscillations are damped out, and the droplet steadily oscillates with the applied frequency  (with time period $t_1=1/\omega=40$)) at an  equilibrium (figure \ref{fig:com_damping}) position $z=z_{shift}$ below the center of the trap. The $z_{shift}$ is responsible for equilibrium COM  oscillation. Such oscillations are not observed if the gravitational force is balanced by some other external force (such as a DC force as used in the experiments \cite{duft03}. For the given experimental parameters, the damping of secular oscillations occurs over a non-dimensional timescale, $t_2=1/(c_z\omega)\sim 7000$, to attain steady equilibrium oscillations. The three timescales in the COM oscillations are indicated in figure \ref{fig:com_damping}. This shows that it is computationally not feasible to carry out numerical simulations for such a long time over which steady-state COM oscillations are observed, even when the viscous corrections are included in the potential flow formalism. To overcome this limitation, the COM oscillations obtained from potential BEM are compared with the solution of modified Mathieu equation (equation \ref{eqn:mathieu_break}) and the two solutions are found to overlap with each other (shown in figure \ref{fig:val_mathieu}). Taking advantage of this, to nullify the effect of secular frequency and remove the initial transience in COM oscillations, we obtained the equilibrium position of the droplet from the solution of equation \ref{eqn:mathieu_break} and the BEM simulations are then carried out with the maximum shifted position (near the south end-cap) of the droplet as an initial condition. With this strategy, COM oscillations of the drop at steady state are obtained, which follows applied frequency and exhibits a $\pi$ phase shift with the applied AC cycle (shown in figure \ref{fig:com_ddvstime}a). 

\begin{figure}
	\begin{center}	
		\includegraphics[width=0.48\textwidth]{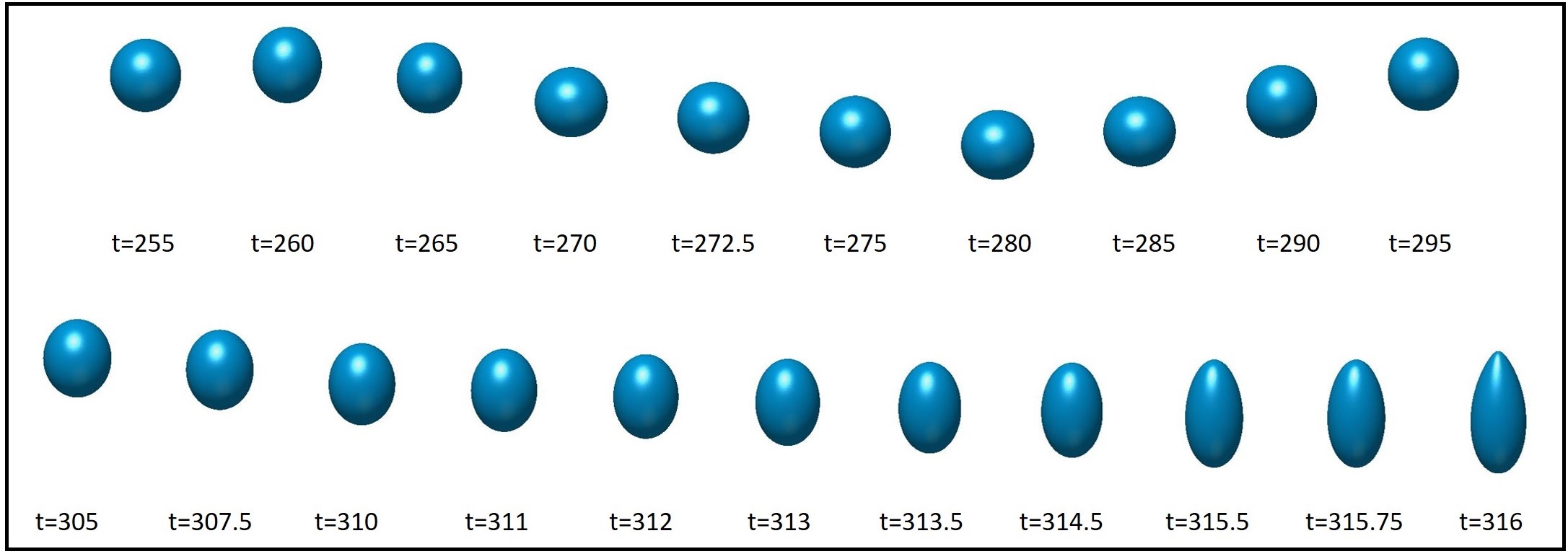}
		\caption{Sequence of drop shapes as a function of time in one cycle of oscillations before the breakup indicating COM motion as well as progressive surface deformation leading to upward breakup. Parameters used are:$Ca_\Lambda=0.0006$, $f=0.025$ and $Bo=0.0038$ and $Q=7.7\pi$.}
		\label{fig:osci_sequence}
	\end{center}
\end{figure}

\begin{figure}[tb]
	\begin{center}	
		\includegraphics[width=0.44\textwidth]{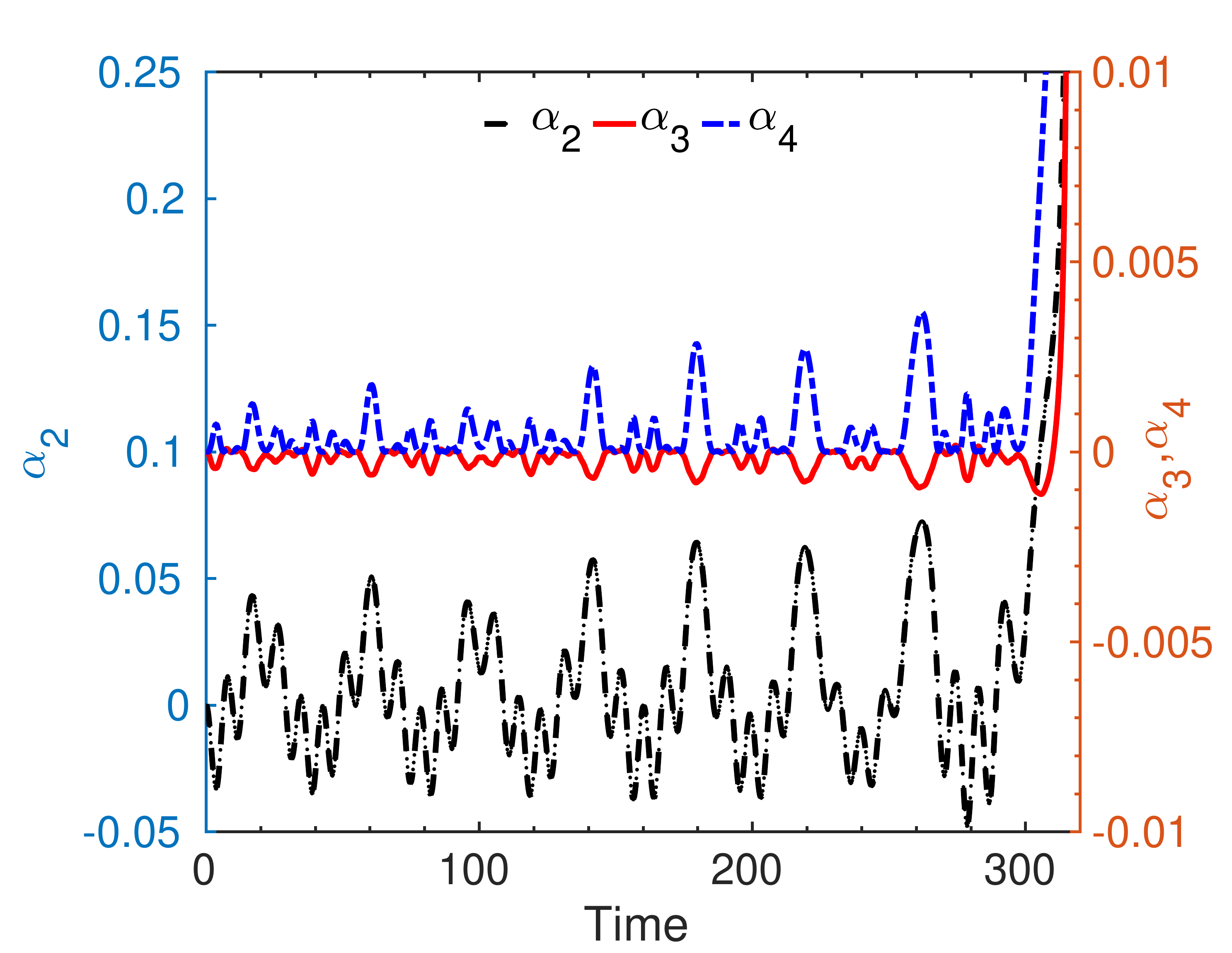}
		\caption{Temporal evolution of amplitudes of $P_2$, $P_3$ and $P_4$ shape modes obtained by decomposing the drop shapes with charge, $Q=7.7\pi$. Note that the simulations are initiated with spherical drop shape.}
		\label{fig:pertubations}
	\end{center}
\end{figure}

The parameters used in the simulations are borrowed from the experiments with charge as a fitting parameter. The simulations are carried out at various values of charge, and the charge at which droplet loses its stability is considered to be the critical charge for the given parameters. For typical experimental parameters of $Ca_\Lambda=0.0006$, $f=0.025$ and $Bo=0.0038$, the droplet breaks at $Q=7.7\pi$ in the upward direction, which is 96.25\% of the Rayleigh limit. The sequence of drop deformation in one oscillation cycle leading to breakup is presented in figure \ref{fig:osci_sequence}. The figure also shows the temporal variation in the relative position of the droplet, indicating COM oscillations similar to that observed in the experiments (\cite{pre_paper}). Figure \ref{fig:com_ddvstime}b shows the degree of deformation as a function of non-dimensional time for $Q=7.7\pi$ indicating that the droplet undergoes few cycles of oscillations developing finite-amplitude perturbation with time that leads to sub-critical instability and eventually breaks in the positive AC cycle at $t=316$ non-dimensional time. It should be noted that the droplet is at the position near the south end cap (at a position= -5.2 in figure \ref{fig:com_ddvstime}a) at $t=0$ when the end caps are at a positive potential and is near the center of the quadrupole trap when the end caps are negative due to $\pi$ phase shift. As the droplet  undergoes COM and surface oscillations, the amplitude of the degree of deformation is seen to increase. This increase is due to a coupling of charge distribution and the applied qaudrupolar AC field. It can be clearly observed in figure \ref{fig:com_ddvstime}b that the critical perturbation (amplitude of degree of deformation) is attained at the negative peak of the AC cycle (at about $t=300$) when the drop position is near the center of the quadrupole field. At this point, the droplet ceases to respond to the applied field, and the further deformation is dominated by charge distribution on the deformed drop. At this point, it appears that with this shape perturbation, the droplet would have undergone a symmetric subcritical, final amplitude instability. However, while the deformation continues to increase, the drop starts shifting towards the south end cap (figure \ref{fig:com_ddvstime}a), in a positive cycle of the AC field. Hence, a positively charged drop near the south end cap experiences higher repulsion at the south pole of the drop from the end cap at a positive potential causing higher charge migration towards the north pole of the already deformed drop, and the drop breaks in the upward direction.     

In order to understand the effect of shape mode coupling on the breakup behavior, the drop shape given by, $r_s(\theta,t)=1+\alpha_l(t) P_l(\cos \theta)$ is decomposed into its linear modes (defined in terms of Legendre polynomials, $P_l$) and the perturbation coefficients ($\alpha_l(t)$) of various modes are obtained using orthogonality condition of Legendre polynomials \cite{lundgren1988oscillations} which gives,
\begin{equation}
\alpha_l(t)=(l+\frac{1}{2})\int_{0}^{\pi}(r_s(\theta,t)-1)P_l(cos\theta)d\theta
\end{equation}
The amplitudes of three linear modes ($P_2$, $P_3$, and $P_4$) as a function of time are shown in figure \ref{fig:pertubations}. The amplitude of the second-mode, $\alpha_2$ is observed to be significantly greater than $\alpha_3$ and $\alpha_4$. This reconfirms that the $P_2$ mode is dominant in the deformation of a charged drop. Although the amplitude of third-mode ($\alpha_3$) is an order of magnitude smaller than $\alpha_2$, it increases as the drop approaches breakup, causing an asymmetric deformation. These results also validate the late onset of asymmetry observed in experiments and numerical analysis in the Stokes flow limit \cite{pre_paper}. The positive value of $\alpha_3$ indicates higher curvature at the north pole and is considered to be responsible for the upward breakup of the drop.     

\begin{figure}[tb]
	\begin{center}	
		\includegraphics[width=0.35\textwidth]{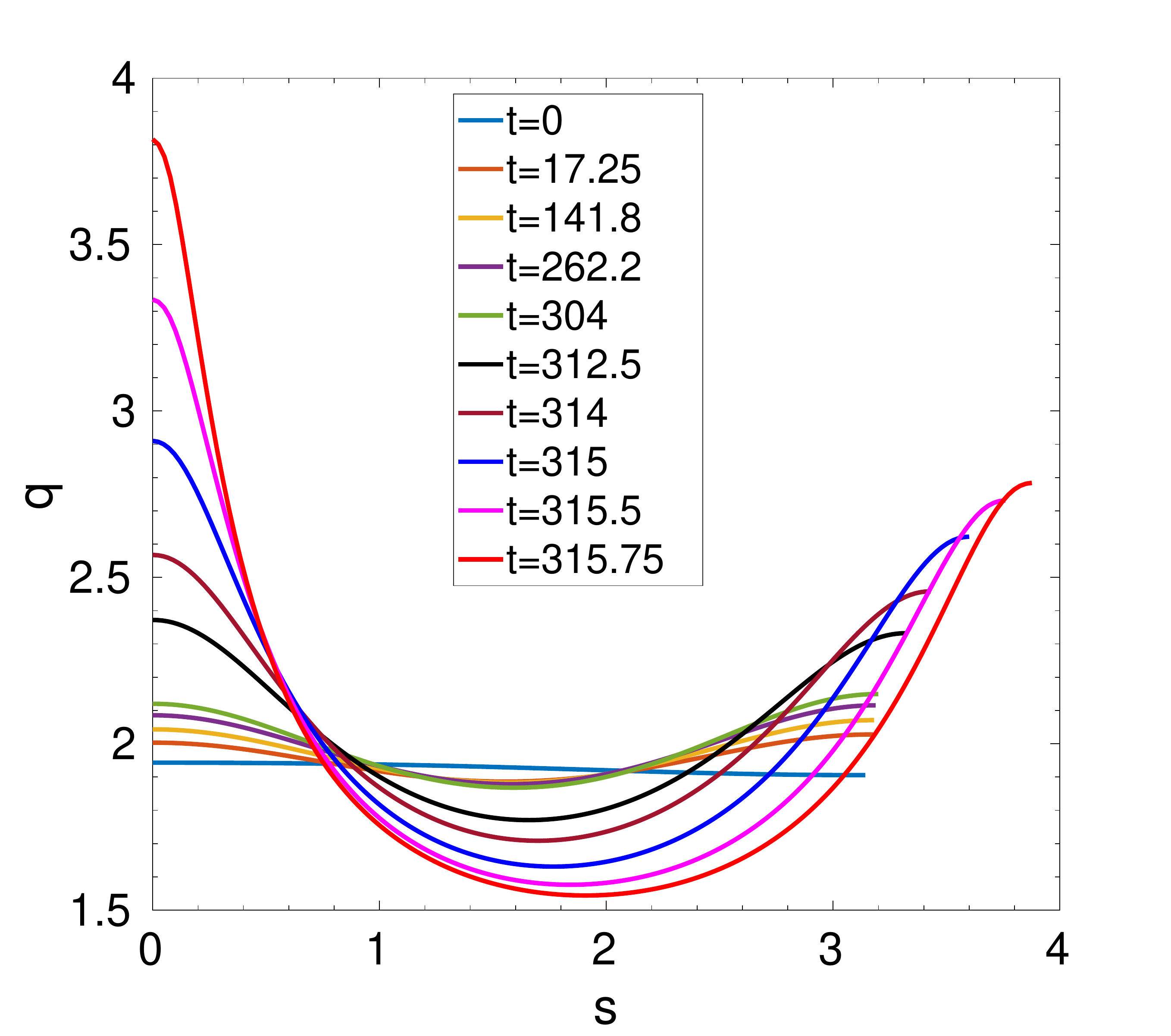}
		\caption{Surface charge density as a function of arclength of the drop for various times when the droplet is near the center of the quadrupole trap (i.e. at the time when DD of the drop is maximum observed in the negative cycle of the applied field). Initially charge density is higher at the south pole of the drop while as the drop deforms the charge density of the north pole increases leading to upward breakup.}
		\label{fig:qVSs}
	\end{center}
\end{figure}

\begin{figure}
	\begin{center}
		\begin{subfigure}[b]{0.49\linewidth}
			\includegraphics[width=\linewidth]{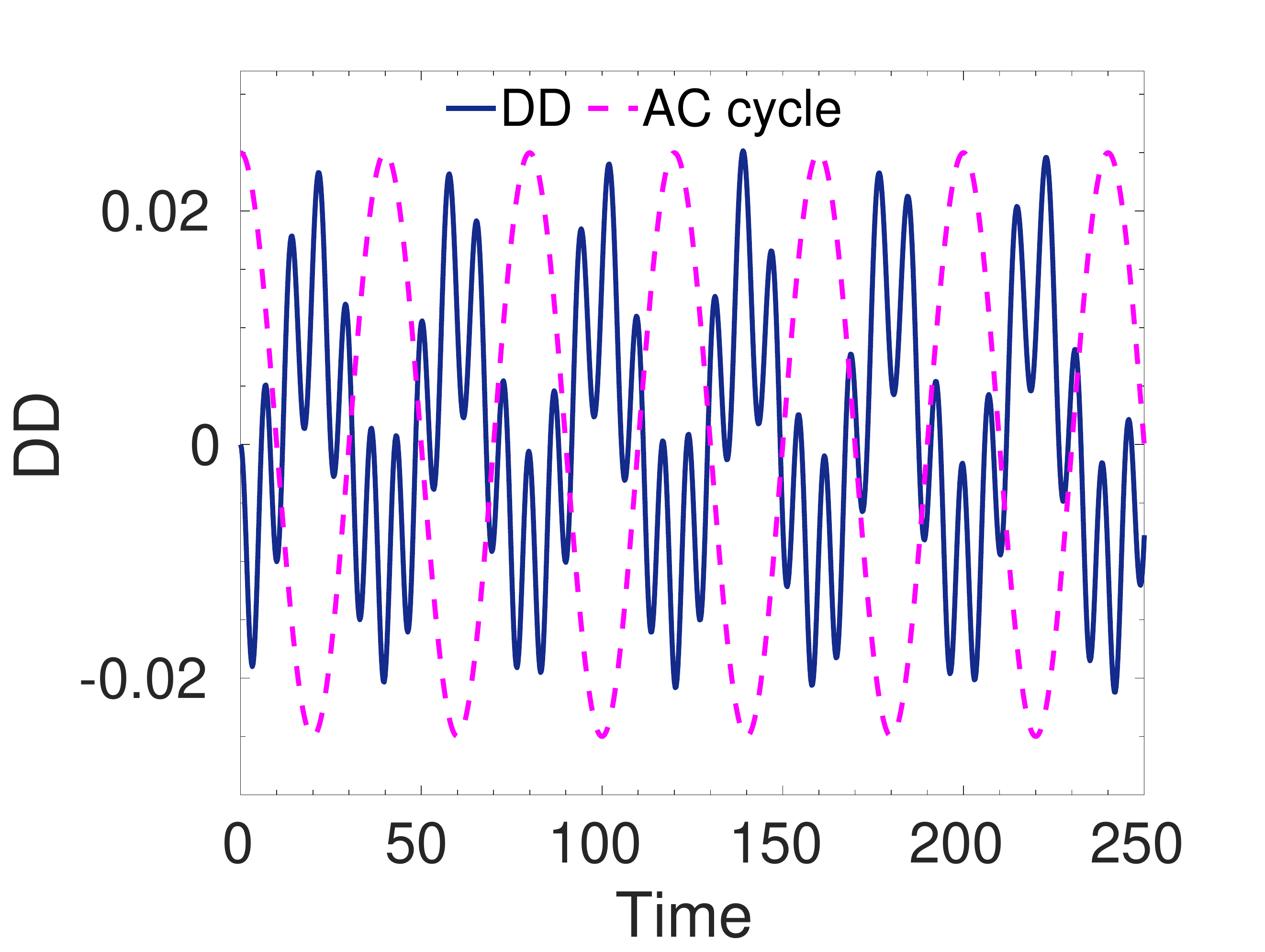}
			\caption{}
			\label{fig:dd7p65}
		\end{subfigure}
		\begin{subfigure}[b]{0.48\linewidth}
			\includegraphics[width=\linewidth]{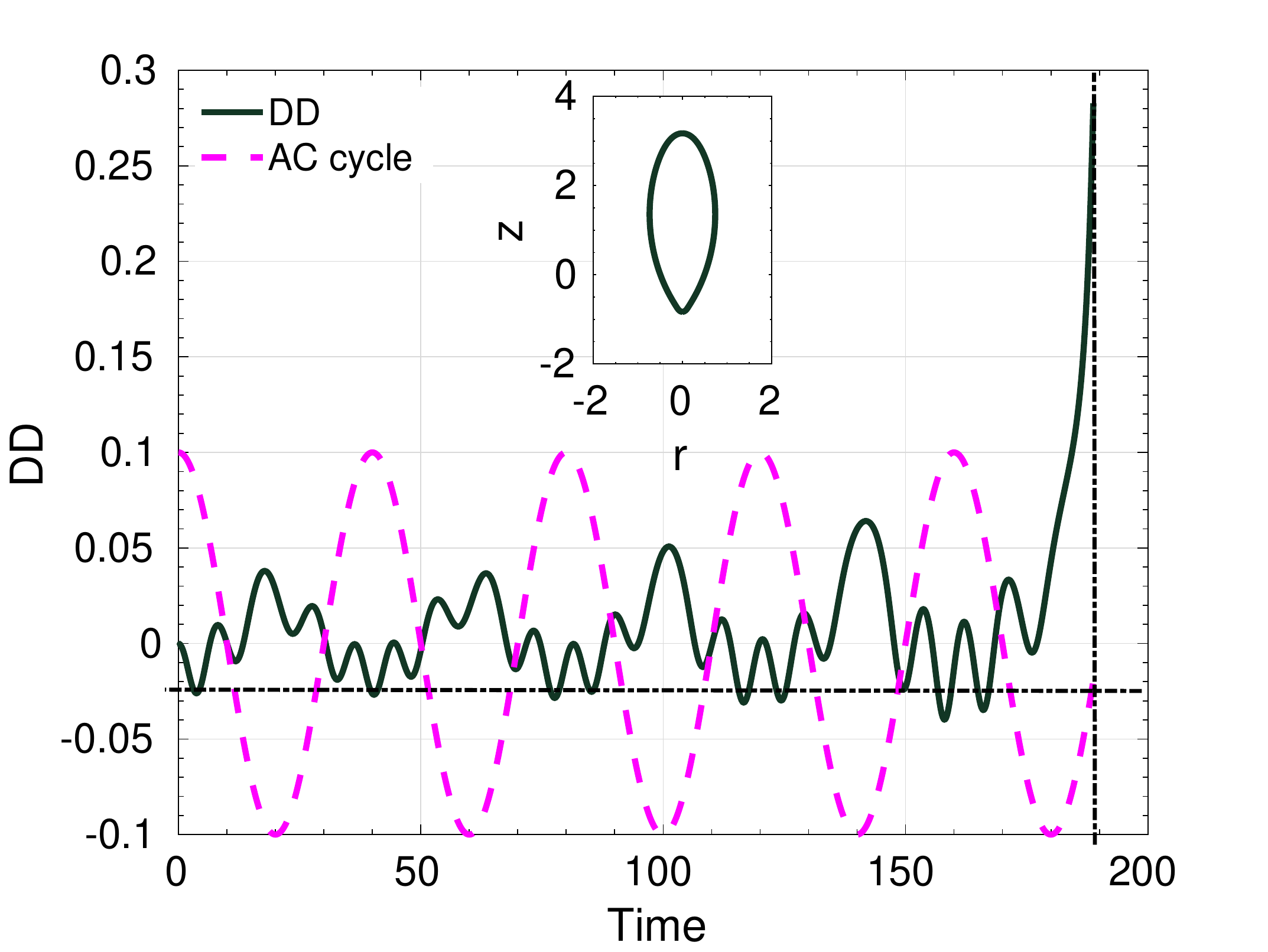}
			\caption{}		
			\label{fig:dd7p72}
		\end{subfigure}
		\caption{Effect of charge on the degree of deformation of the drop at $Ca_\Lambda=0.0006$, $Bo=0.0038$ and $f=0.025$(a)Stable oscillations till $t=250$ for $Q=7.65\pi$ and (b)downward drop breakup at $Q=7.72\pi$.}
	\end{center}
\end{figure}
Further, to understand the dynamics of the process, we analyzed the charge density distribution on the drop surface at the time when the deformation curve shows maximum prolate deformation. The charge density as a function of arclength is shown in figure \ref{fig:qVSs} for various times. Figure \ref{fig:qVSs} indicates that initially when the droplet is undergoing oscillations with small deformation (arclength ($s$) is almost constant), the charge density at the south pole is marginally higher as compared to the north pole of the drop. This is mainly due to the fact that the positively charged droplet is shifted from the center in the downward direction and thus experiences greater attraction from the south end cap in the negative AC cycle, resulting in greater accumulation of charges at the south pole of the drop. However, as the drop starts deforming (arclength of the drop increases) with time, the charge density distribution becomes symmetric about the equator of the drop with equal charge densities at the north and the south poles. Further, when the drop attains a critical deformation, the shape coupling with charge distribution becomes dominant, and the charge density is found to increase at the north pole (figure \ref{fig:qVSs}(a)) due to higher repulsion from the south end cap. This leads to the accumulation of charges in the north pole, which in turn causes the droplet to break in the upward direction.
\begin{figure}
	\begin{center}
		\begin{subfigure}[b]{1\linewidth}
			\includegraphics[width=\linewidth]{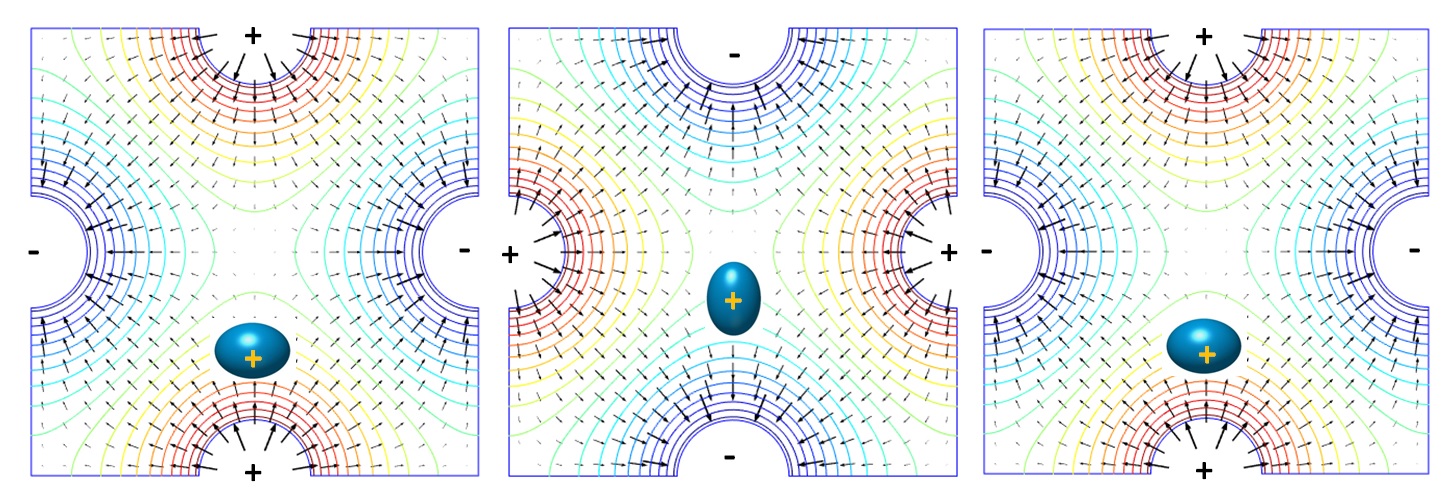}
			\caption{}
			\label{fig:low}
		\end{subfigure}
		\begin{subfigure}[b]{1\linewidth}
			\includegraphics[width=\linewidth]{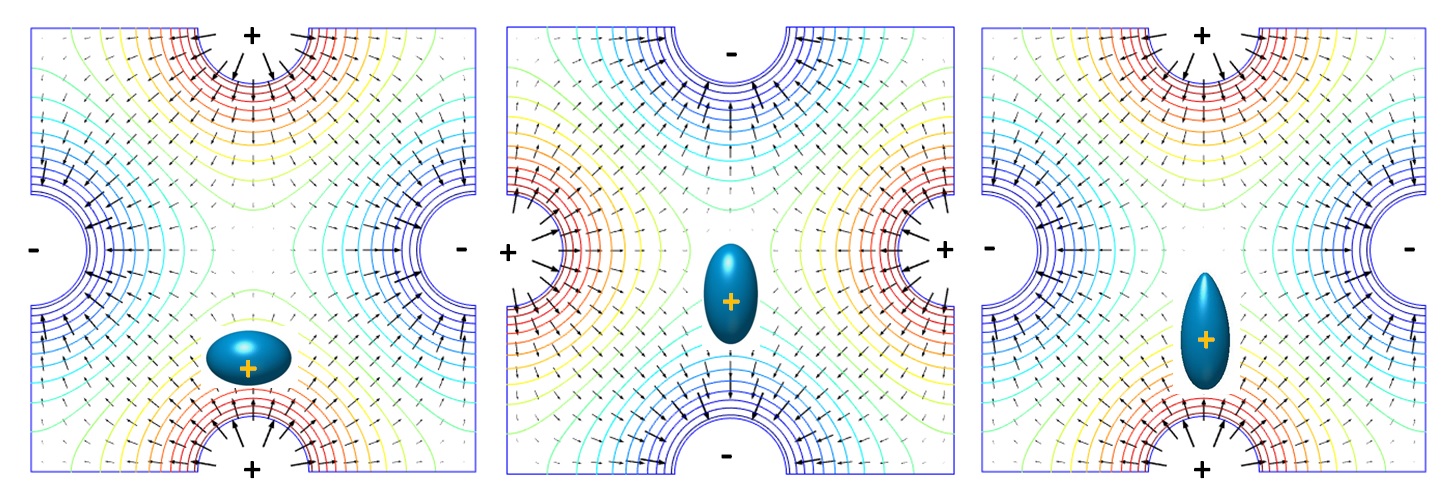}
			\caption{}		
			\label{fig:critical}
		\end{subfigure}
		\begin{subfigure}[b]{1\linewidth}
			\includegraphics[width=\linewidth]{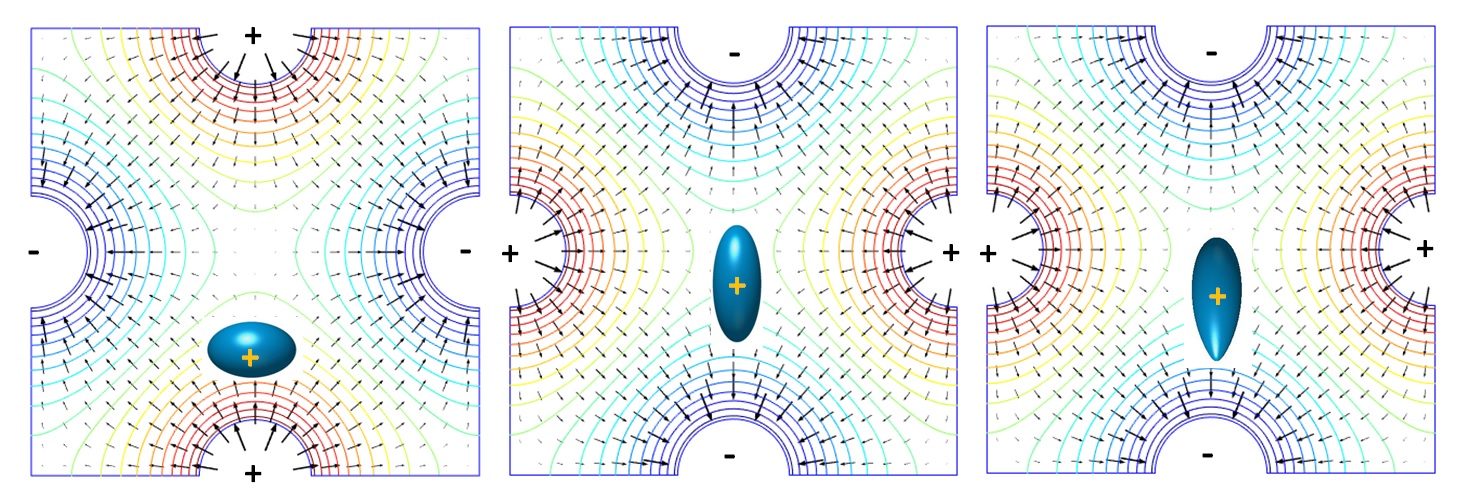}
			\caption{}		
			\label{fig:high}
		\end{subfigure}
		\caption{Illustration of sub-critical asymmetric breakup of a charged drop in quadrupole field for charge, (a)lower ($Q\sim7.65\pi$), (b) critical ($Q=7.7\pi$) and (c) high ($Q\sim7.72\pi$) at constant $Ca_\Lambda$ and $Bo$.}
		\label{fig:breakup_scheme}
	\end{center}
\end{figure}

At a lower value of charge, $Q<7.7 \pi (~7.65\pi)$,  the drop exhibits stable oscillations for a long time without showing any signature of increase in the degree of deformation (simulations are carried out till $t=250$ as shown in figure \ref{fig:dd7p65}). This shows that, at $Q=7.65\pi$, the drop cannot attain sufficient deformation to become Coulombially unstable, and thus it continues to respond to the applied AC field and exhibit stable, sustained oscillations. On the other hand, for $Q>7.7 \pi(~7.72\pi)$, the drop breaks in the downward direction (figure \ref{fig:dd7p72}). This indicates that  a critical deformation is achieved at the negative peak potential of end caps, which is sufficient to induce Rayleigh instability in the drop. It should be noted that, as the droplet considered is positively charged, it attains the highest prolate perturbation when the end caps have negative polarity. In this situation, due to the presence of a high surface charge, the south pole of the drop experiences higher attraction, and the droplet breaks in the downward direction before the AC cycle (and thereby the polarity of the end  cap) changes (figure \ref{fig:dd7p72}). These results re-confirm the hypothesis given for explaining the directionality of breakup of the charged drop using viscous BEM simulations discussed in our previous work \cite{pre_paper}. In this work it is shown that the numerical simulations in potential flow limit simultaneously capture the oscillation phase, including both coupled COM and surface oscillations, progressive development of critical deformation and its subsequent breakup all in one simulation. 

To recapitulate, the experimental observation of asymmetric breakup of a levitated charged droplet is explained in detail using figure \ref{fig:breakup_scheme}. Figure \ref{fig:breakup_scheme} (a) indicates that at a lower charge, the droplet oscillates without breakup, exhibiting stable shape oscillations responding to the applied AC field. As the time progresses, the total charge density on the drop increases due to evaporation, and it undergoes large amplitude deformations. As soon as the value of charge reaches to a critical value (which is less than the Rayleigh limit) an instability sets in, leading to droplet breakup, clearly demonstrating the subcriticality of the instability (figure \ref{fig:breakup_scheme} (b)). At this stage, if the droplet is positively charged and attains the critical deformation in the positive cycle of the applied AC field, the asymmetry is rendered in such a way that the droplet breaks in the upward direction. However, if the critical conditions with respect to deformation and charge on the droplet are attained when the droplet is near the end cap whose polarity is still negative, the symmetric subcritical instability is rendered asymmetric with downward direction of breakup (figure \ref{fig:breakup_scheme} (c)). 

The BEM calculations in the potential flow limit not only capture several cycles of coupled COM motion and surface oscillations of a levitated charged drop before the breakup but also show the simultaneous build-up of critical perturbation that is observed to be responsible for the sub-critical asymmetric Rayleigh breakup. The study shows that in the presence of external fields there is a strong coupling in the dynamics of COM and shape of the charged drops. The numerical simulations are also extended to check the effect of strength of the electric field which indicate that the asymmetry in drop shape at the onset of breakup increases with the strength of the applied quadrupole field and the positional shift of droplet from the center of the field. Thus it can be conjectured that, at the electric field strengths required to levitate the droplet, the external fields act as initiators of finite-amplitude shape deformations, that only influence the Rayleigh breakup process while the induced instability can be seen at much higher values. 

Detailed numerical simulations that simultaneously describe the electrodynamic levitation, deformation, and breakup of a charged drop in a quadrupole field are presented for the first time. While the work of Duft \etal \cite{duft03} has shown pioneering evidence for a symmetrical pathway for Rayleigh breakup, the present study shows that this is not universal, and perhaps is an exception in realistic situations. Specifically, by considering the effect of gravity and external electric fields, an asymmetric breakup might turn out to be the rule in real-life practical situations, such as in nano-drop generators using electrosprays. The study presented here, also has far-reaching practical implications on the ion mass spectrometry which depends on the basic principle of Rayleigh break up of charged droplets. The occurrence of sub-criticality of the instability and its dependence on the polarity and strength of the confining electrodes will not only influence the time line of the process but also modify the size distribution of droplets thus formed. Hence it is necessary to carefully consider the findings presented here in the future studies.

\acknowledgments
Authors would like to acknowledge Dr. Mohit Singh for his valuable discussions borne from his experimental observations. This work was funded by the Bhabha Atomic Research Centre (BARC) of DAE under Grant No. 36(4)/14/32/2014-BRNS/1406.


\begin{thebibliography}{0}
  
\bibitem{rayleigh1882}
  \Name{Rayleigh, L.}
  \REVIEW{The London, Edinburgh, and Dublin Philosophical Magazine and Journal of Science}{14}{184}{1882}.
  
\bibitem{duft03}
\Name{Duft D., Achtzehn T., M\"{u}ller R., Huber B.A. \and Leisner T.},
\REVIEW{Nature}{421}{128}{2003}.

\bibitem{duft02}
  \Name{Duft D., Lebius H., Huber B.A., Guet C.\and Leisner T.}
  \REVIEW{Physical review letters}{89}{084503}{2002}.


\bibitem{giglio08}
  \Name {Giglio E., Gervais B., Rangama J., Manil B., Huber B.A., Duft D., M\"{u}ller R., Leisner T. \and Guet C.}
  \REVIEW{Physical Review E}{77}{036319}{2008}.
 
\bibitem{feng1991three}
  \Name{Feng J.Q. \and Beard K.V.,}
  \REVIEW{Journal of fluid mechanics}{227}{429}{1991}.

\bibitem{singh2019effect}
  \Name{Singh M., Gawande N., Mayya Y.S. \and Thaokar R.}
  \REVIEW{Langmuir}{35}{15759}{2019}.

\bibitem{adornato83}
  \Name{Adornato P.M. \and Brown R.A}
  \REVIEW{Proceedings of the Royal Society of London A: Mathematical, Physical and Engineering Sciences}{389}{1796}{1983}.

\bibitem{feng1997}
  \Name{Feng Z.C.}
  \REVIEW{Journal of Fluid Mechanics}{333}{1}{1997}.

\bibitem{natarajan1987}
  \Name{Natarajan R. \and Brown R.A.}
  \REVIEW{Proc. R. Soc. Lond. A}{410}{209}{1987}.

\bibitem{pelekasis1990equilibrium}
  \Name{Pelekasis N.A., Tsamopoulos J.A. \and Manolis G.D.}
  \REVIEW{Physics of Fluids A: Fluid Dynamics}{2}{1328}{1990}

\bibitem{tsamopoulos85}
  \Name{Tsamopoulos J.A., Akylas T.R. \and Brown R.A.}
  \REVIEW{Proceedings of the Royal Society of London A: Mathematical, Physical and Engineering Sciences}{401}{6}{1985}.

\bibitem{basaran1989}
  \Name{Basaran O.A. \and Scriven L.E.}
  \REVIEW{Physics of Fluids A: Fluid Dynamics}{1}{799}{1989}.

\bibitem{pelekasis1995dynamics}
  \Name{Pelekasis N.A. \and Tsamopoulos J.A.}
  \REVIEW{Engineering analysis with boundary elements}{15}{339}{1995}.

\bibitem{das15}
  \Name{Das S., Mayya Y.S. \and Thaokar R.}
  \REVIEW{EPL (Europhysics Letters)}{111}{24006}{2015}.


\bibitem{singh2018}
	\Name{Singh M., Thaokar R., Khan A. \and Mayya Y.S.}
	\REVIEW{Physical Review E}{98}{032202}{2018}.

\bibitem{lundgren1988oscillations}
  \Name{Lundgren T.S. \and Mansour N.N.}
  \REVIEW{Journal of Fluid Mechanics}{194}{479}{1988}.

\bibitem{feng1996numerical}
  \Name{Feng Z.C. \and Leal L.G.}
  \REVIEW{International Journal of Multiphase Flow}{22}{93}{1996}.

\bibitem{baker1982generalized}
	\Name{Baker G. R., Meiron D. I. \and Orszag S. A.}
	\REVIEW{Journal of Fluid Mechanics}{123}{477}{1982}.

\bibitem{singh2020influence}
  \Name{Singh M., Gawande N. \and Thaokar R.}
  \REVIEW{Journal of Applied Physics}{128}{145304}{2020}.
  
\bibitem{betelu2006}
\Name{Betelú S.I., Fontelos M.A., Kindelán U. \and Vantzos O.}
\REVIEW{Physics of Fluids}{18}{051706}{2006}.

\bibitem{gawande2017}
\Name{Gawande N., Mayya Y.S. \and Thaokar R.}
\REVIEW{Physical Review Fluids}{2}{113603}{2017}.

\bibitem{gawande2020}
\Name{Gawande N., Mayya Y.S. \and Thaokar R.}
\REVIEW{Journal of Fluid Mechanics}{884}{2020}.

\bibitem{pre_paper}
\Name{Singh M., Gawande N., MAyya Y. S., Thaokar R.}
\REVIEW{Physical Review E}{103}{053111}{2021}

\bibitem{singh2018pof}
\Name{Singh, M., Gawande, N., Mayya, Y.S. \and Thaokar, R.}
\REVIEW{Physics of Fluids}{30(12)}{122105}{2018}.

\end{thebibliography}
\end{document}